\documentclass{vgtc}                          

\graphicspath{{figures/}{pictures/}{images/}{./}} 

\usepackage{times}                     

\usepackage{tabu}                      
\usepackage{booktabs}                  
\usepackage{lipsum}                    
\usepackage{mwe}                       

\usepackage[tagged, highstructure]{accessibility} 

\newcommand{\systemname}{\emph{ReVise}\xspace}

\usepackage{mathptmx}                  

\onlineid{1200}

\vgtccategory{Research}

\vgtcinsertpkg

\title{\textit{ReVise}: A Human-AI Interface for Incremental Algorithmic Recourse}

\author{Kaustav Bhattacharjee\thanks{e-mail: kaustav.bhattacharjee@pnnl.gov}\\ %
        \parbox{2in}{\scriptsize \centering Optimization \& Control Group \\ Pacific Northwest National Laboratory} %
\and Jun Yuan\thanks{e-mail: jy448@njit.edu}\\ %
     \parbox{2in}{\scriptsize \centering Department of Data Science \\ New Jersey Institute of Technology} %
\and Aritra Dasgupta\thanks{e-mail: aritra.dasgupta@njit.edu}\\ %
     \parbox{2in}{\scriptsize \centering Department of Data Science \\ New Jersey Institute of Technology}
     }

\abstract{
    The recent adoption of artificial intelligence in socio-technical systems raises concerns about the black-box nature of the resulting decisions in fields such as hiring, finance, admissions, etc.
If data subjects—such as job applicants, loan applicants, and students—receive an unfavorable outcome, they may be interested in algorithmic recourse, which involves updating certain features to yield a more favorable result when re-evaluated by algorithmic decision-making.
Unfortunately, when individuals do not fully understand the incremental steps needed to change their circumstances, they risk following misguided paths that can lead to significant, long-term adverse consequences.
Existing recourse approaches focus exclusively on the \textit{final} recourse goal but neglect the possible incremental steps to reach the goal with real-life constraints, user preferences, and model artifacts.
To address this gap, we formulate a visual analytic workflow for \textit{incremental} recourse planning in collaboration with AI/ML experts and contribute an interactive visualization interface that helps data subjects efficiently navigate the recourse alternatives and make an informed decision. We present a usage scenario and subjective feedback from observational studies with twelve graduate students using a real-world dataset, which demonstrates that our approach can be instrumental for data subjects in choosing a suitable recourse path.
} 


\begin{document}


\firstsection{Introduction}

\maketitle
Artificial intelligence (AI) drives algorithmic decision-making in various socio-technical contexts such as hiring, lending, and admissions, where individuals — serving as data subjects — experience tangible impacts based on the outcomes produced by these systems. Data subjects who receive undesired outcomes may seek algorithmic recourse, i.e., the process of updating certain features to yield improved outcomes when re-evaluated by the decision-making process~\cite{guo2020survey, poyiadzi2020face}. Yet, achieving effective recourse is complex; it must account for the data subject's initial feature values, their unique preferences and real-life constraints, the decision-making rules of the model, and the influence of other data subjects processed by the model. Two main forms of recourse exist in the literature: contrastive explanation~\cite{millerContrastiveExplanationStructuralModel2021}, which offers examples of others who received favorable outcomes, and consequential recommendations~\cite{ustunActionableRecourseLinear2019}, which suggest actions to achieve favorable outcomes in the future. However, both approaches focus predominantly on the final outcome rather than on the step-by-step process of reaching recourse. This emphasis on the ultimate goal, rather than on the incremental journey, poses challenges in practical, real-world scenarios where gradual adjustments are often necessary.

Our review of the literature on algorithmic recourse reveals two principal limitations in current approaches. First, many methods treat recourse as a deterministic problem aimed solely at achieving a specific favorable outcome, often neglecting the incremental progression from a negative state. For example, a loan applicant might be instructed to boost their credit score by a fixed amount to secure approval, without considering that incremental improvements could gradually enhance their likelihood of success. Second, existing frameworks tend to prescribe changes in feature values without allowing individuals the flexibility to select tailored actions along their improvement journey. One might imagine a job applicant being advised to pursue a single, rigid upgrade—such as obtaining another degree—instead of exploring a series of manageable steps that align with their personal constraints and preferences. This narrow focus on binary endpoints disregards the inherent uncertainties of real-world settings, where understanding gradual paths is critical. To address these problems, we introduce \textit{incremental recourse}, guiding data subjects through a sequence of steps toward desired outcomes rather than presenting a single change. Instead of focusing on binary outcomes, we utilize probability scores to show gradual progress—for example, using the probability of loan approval rather than simply approval/disapproval designations.

We implement a human-in-the-loop approach for incremental recourse planning where users actively participate in defining their recourse path. Users update their data to define subsequent recourse states based on their preferences, real-life constraints, and insights from other data subjects' experiences. This approach requires understanding model behavior locally, which presents challenges when working with black-box models. We incorporate feature attribution explanations in our interactive workflow as they represent local model preferences, allowing users to make informed choices about their next steps rather than randomly testing for better outcomes.
We contribute \systemname, an interactive recourse planning assistant, which facilitates multi-way comparison and decision-making among recourse states and paths, supported by interactive visualization designs for human-in-the-loop recourse calibration.
We focus specifically on decision-making tasks that underpin the visualization design space, moving beyond the sense-making or exploration tasks commonly addressed in visualization research~\cite{dimaraCriticalReflectionVisualization2022}.
Our visual analytic approach enables users to compare alternative paths efficiently, understand trade-offs between different feature changes, and make informed decisions about their recourse journey with greater agency and awareness of incremental probability improvements.

\section{Related Work}
Algorithmic recourse~\cite{karimiSurveyAlgorithmicRecourse2023} generally follows two main approaches: contrastive explanation~\cite{millerContrastiveExplanationStructuralModel2021} (``applicants with credit scores above $720$ were approved") and consequential recommendations~\cite{ustunActionableRecourseLinear2019} (``increase your credit score by $50$ points to improve approval chances"). Our focus on \textit{incremental} recourse differs from both by emphasizing momentum toward favorable outcomes rather than guaranteeing them, constructing action sequences where features may be improved multiple times. Karimi et al. argue that structured causal models are essential for improving the feasibility of recourse, distinguishing between mere explanations and meaningful interventions for actionable change~\cite{karimiAlgorithmicRecourseCounterfactual2021}. However, developing precise causal models is challenging and often influenced by individual perspectives, which can lead automated recourse to produce unrealistic recommendations. In this work, we offer visual interpretations of actionable incremental targets, enabling users to adjust and customize the machine-generated suggestions according to their own preferences.

Feature attribution methods quantify the contribution of each feature to a model’s prediction, an insight vital for algorithmic recourse as it highlights which feature modifications most effectively change outcomes. Researchers have linked feature attribution with recourse via counterfactual explanations—unifying attribution and counterfactuals~\cite{kommiyamothilalUnifyingFeatureAttribution2021} and creating counterfactual Shapley additive explanations~\cite{albiniCounterfactualShapleyAdditive2022}. Additional approaches have leveraged techniques such as: generating counterfactual and contrastive explanations using SHAP~\cite{rathiGeneratingCounterfactualContrastive2019}; employing Shapley values to estimate the combinatorial effect of multiple changes to generate counterfactual text~\cite{fernTextCounterfactualsLatent2021}; and proposing methods to derive counterfactual explanations using feature attributions generated by LIME and SHAP~\cite{ramonComparisonInstancelevelCounterfactual2020}.
However, some critiques argue that using attribution-based explanations as final recourse suggestions may not be robust enough~\cite{fokkemaAttributionbasedExplanationsThat} or require reformulation for greater intuitiveness~\cite{fernandez-loriaExplainingDataDrivenDecisions2021}. In this work, we utilize feature attribution to form visual cues for incremental counterfactual suggestions by limiting explanations to local regions, which facilitates a practical assessment of actionability and forms the foundation of \systemname.

Existing visualization-based interfaces for recourse include DECE, which uses contrastive examples within user-defined feature ranges for neural network predictions~\cite{cheng2020dece}; GAM Coach’s text-heavy, tab-based interface~\cite{wang2023gam}; and tools such as the what-if tool~\cite{wexlerWhatIfToolInteractive2019} and ViCE~\cite{gomezViCEVisualCounterfactual2020}. GAM Coach’s reliance on tabbed navigation hinders multi-way comparisons and offers limited support for visualizing progressive recourse trajectories, and none of these systems explicitly facilitate decision-making tasks—investigating, synthesizing, and selecting among alternatives—which are crucial for incremental recourse planning~\cite{dimaraCriticalReflectionVisualization2022}. To the best of our knowledge, our work is the first in visual analytics to bridge the gap between incremental recourse needs, the corresponding metrics, and the interactive visualizations required to effectively communicate them to end users.
\section{Visual Analytic Tasks \& Interface Design}

The design space for developing a visualization system for incremental recourse planning is driven by the need for a transparent explanation of the recourse steps and the clear evaluation of potential next steps. In this section, we first discuss the visual analytic tasks we formulated to achieve these goals in our web-based interface, and then explore the supporting design elements.

\subsection{Visual Analytic Tasks}
We have identified six key visual analytic tasks that form the basis of the design space for incremental recourse planning—the first three support understanding the current state, while the next three facilitate informed decision-making for the next recourse state.

\par \noindent \textbf{T1}: \textit{Understanding current state features and attributions:} The first step in recourse planning is to understand the current state across various features and their corresponding attributions. This task relates to the understanding and comparison of the feature values for the current state and their attributions.

\par \noindent \textbf{T2}: \textit{Assessing impact of features on model outcomes:} Understanding the model outcome and the impact of each feature on the outcome is an essential step for recourse planning. This task addresses understanding the model outcome in relation to each feature.

\par \noindent \textbf{T3}: \textit{Comparing feature deviation from current state averages}: A user may wish to understand how their current feature values deviate from averages within normalized feature ranges, enabling them to identify which features are above or below typical values and prioritize improvements accordingly.

\par \noindent \textbf{T4}: \textit{Identifying potential targets for next state}: Users may be willing to understand which targets would be suitable for the next state based on the projection values of all the possible targets. This task revolves around aiding the users to find a target for the next state from a pool of candidate targets.

\par \noindent \textbf{T5}: \textit{Comparing potential targets across features}: The potential targets may have different attributions for each feature. Hence, it is essential to compare them across different features to refine the selection of the potential targets.

\par \noindent \textbf{T6}: \textit{Comparing recourse paths}: Selecting a target for the next state creates a recourse path, which can be incrementally modified to reach the desired target. This task relates to the iterative modification of the recourse paths by comparing the model outcomes and feature attributions through different recourse states.
\begin{figure*}
    \includegraphics[width=\textwidth]{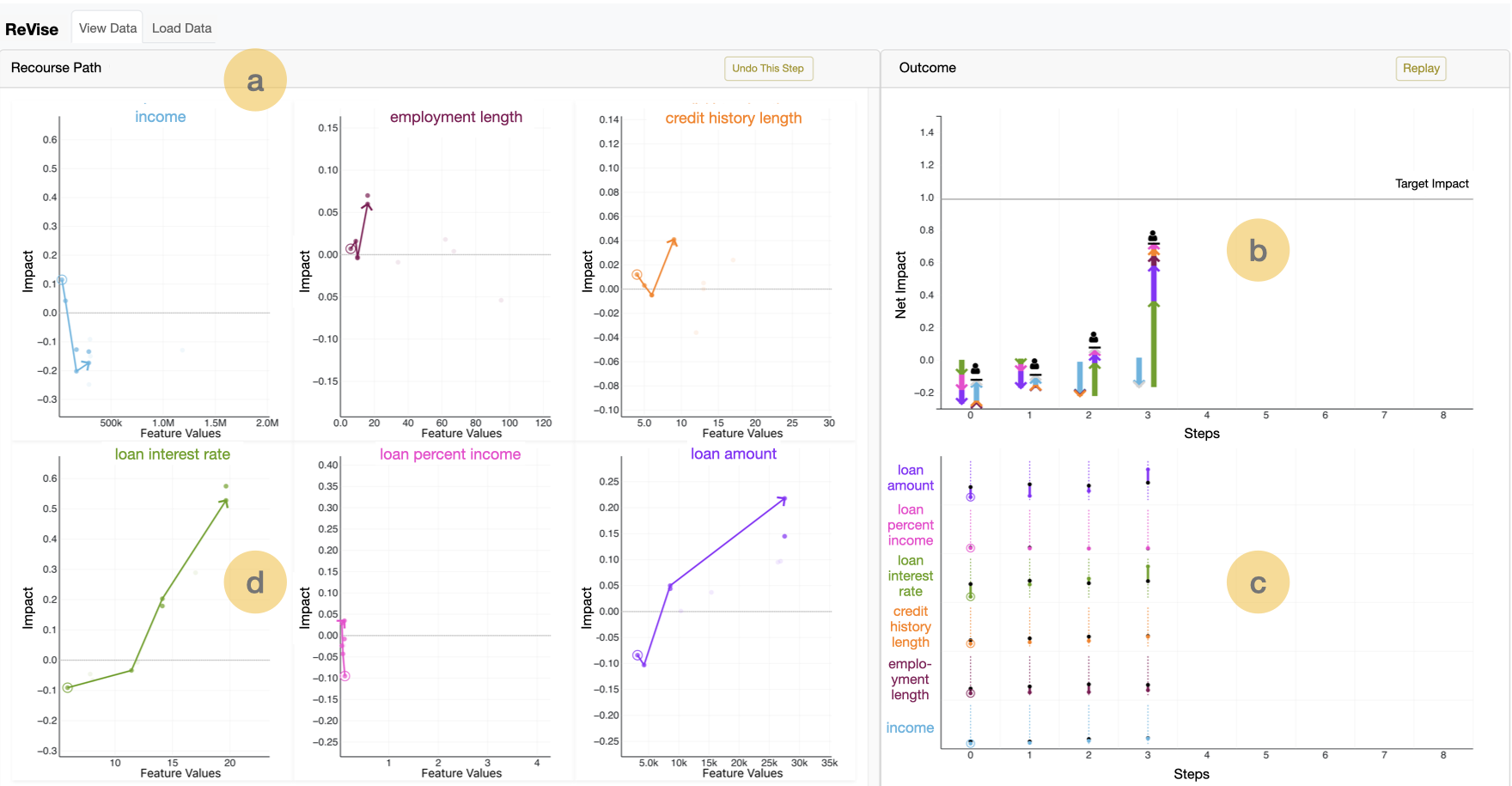}
     \vspace{-4mm}
        \caption{\textbf{Recourse Planning Usage Scenario:} (a) Utilizing the Credit Risk dataset, the user selected a recourse path from the Recourse Path Explorer. (b) After multiple iterations, she identified a recourse path that resulted in significant improvement in the model outcome. (c) At the same time, the user observed that some features reached significantly higher-than-average values, while others showed a downward trend in their attributions. (d) Despite these differences, the majority of features followed an upward trajectory, consistent with the overall aim of recourse planning. Ultimately, the user confirmed that \systemname's visual encodings faithfully reflect the explanation results offered by SHAP.
        }
        \alt{Screenshot of the ReVise interface showing a recourse planning usage scenario with multiple feature impact plots for income, employment length, credit history length, loan interest rate, loan percent income, and loan amount, alongside outcome monitors that visualize how feature changes and recourse steps affect the predicted model outcome, illustrating the user’s selection and evaluation of recourse paths and the faithful representation of SHAP-based explanations.}
        \label{fig:caseStudy}
        \vspace{-2mm}
\end{figure*}

\subsection{Visual Cues for Expressing Recourse Patterns}
In this subsection, we align recourse components with design elements, highlighting the cues and visual analytic tasks that facilitate incremental recourse for one or more data subjects.

\par \noindent  \textbf{Cues for Associating Feature Attribution and Value}: The attribution for a data subject's feature is determined and estimated through a feature explanation method (e.g., SHAP~\cite{lundberg2017unified}) which distributes the outcome across all features of the data subject. Each data subject in the current state is depicted as a dot in a scatterplot~(Figure~\ref{fig:caseStudy}a), where the dot's position reflects its feature value and corresponding attribution—for instance, a loan applicant's employment length of $5$ years might have a positive attribution of $0.15$ toward loan approval~\textbf{(T1)}. Hovering over a dot highlights its position relative to the overall current state, enabling comparison of potential targets for the next state~\textbf{(T5)}.

\par \noindent  \textbf{Cues for Part-to-Whole Impact}: Different features have varying impacts on the model outcome for a particular data subject at a certain state. The impact of each feature is encoded by the length of its attribution, also illustrating its contribution relative to the overall outcome~(Figure~\ref{fig:caseStudy}b)~\textbf{(T2)}. For example, this design may reveal that the loan interest rate has a more significant effect than the loan amount for a particular data subject, suggesting that improving the interest rate could yield better outcomes in the next recourse step. To assess feature value deviations at each recourse step~\textbf{(T3)}, we represent each feature using a dotted line that spans the normalized range, with a black dot indicating the average feature value across all available recourse states and a colored dot representing the current feature value~(Figure~\ref{fig:caseStudy}c). A solid colored line connects these dots to visualize the deviation—whether an increase or decrease—from the average, enabling users to quickly identify features that differ significantly from the norm for strategic decision-making about which features to prioritize in the next recourse state.

\par \noindent  \textbf{Cues for Direction}: Another important aspect of understanding a feature's impact on the model outcome is the direction of its attribution, as a data subject may find all features contributing negatively in one state, while in another, some features may contribute positively and others negatively~(Figure~\ref{fig:caseStudy}b)~\textbf{(T2)}. By addressing features with negative impacts—selecting a potential target for the next state with higher attribution for those features—a more favorable recourse path may be achieved. The direction of the recourse trajectory indicates whether the chosen path is leading toward improvement; an upward trajectory signifies progress~(Figure~\ref{fig:caseStudy}d), whereas a downward trend suggests that the selected path may be suboptimal. This encoding allows the data subject to backtrack to a previous state and iterate through a different recourse path based on alternative potential targets~\textbf{(T6)}.

\par \noindent  \textbf{Cues for Actionability}: Selecting the potential target for the next state can be challenging given the pool of candidate targets with varying feature values and attributions. To overcome this, we use a change-to-attribution ratio (also called projection) to identify the top targets for the next state~\textbf{(T4)}, visually encoding it through dot size with the top three targets highlighted by larger dots.
We considered an alternative design that resized all dots based on their projection values, but feedback showed that abstracting to only the top few targets was more effective.

\par \noindent  \textbf{Cues for Trajectory}: The slope of the recourse path indicates the change magnitude; a steeper slope reflects a larger attribution gain per unit feature change, aligning with the ``minimum effort, maximum output" principle—as when a two-year increase in credit history length yields a 70\% attribution rise. Users can iterate through different paths to find steeper, more promising trajectories~\textbf{(T6)}.

\subsection{Interactive Interface}
In this subsection, we introduce the web-based interface for \systemname, designed to facilitate incremental recourse planning for data subjects and collaborating explainable AI experts. The interface comprises two main views: the Recourse Path Explorer~(Figure\ref{fig:caseStudy}a) assists users in initiating recourse planning, understanding their current state, and identifying potential targets~\textbf{(T1}, \textbf{T4}, \textbf{T5}, \textbf{T6)}, while the Outcome Monitor~(Figures~\ref{fig:caseStudy}b and \ref{fig:caseStudy}c) tracks model outcomes with corresponding feature impacts~\textbf{(T2)} and shows deviations from average feature values~\textbf{(T3)}. Developed with React.js and D3.js, this coordinated layout streamlines the exploration of recourse paths while simultaneously assessing their effectiveness.

\subsubsection{Coordinated Multiple Views}

\par \noindent  \textbf{Recourse Path Explorer}: The Recourse Path Explorer presents six scatter plots arranged in two rows to visualize correlations between feature values and attributions for the most influential features, with each dot representing a data subject~(Figure\ref{fig:caseStudy}a). Users can select any dot, causing corresponding dots across all plots to be highlighted; selecting multiple dots creates solid lines connecting them in sequence, with arrows indicating the trajectory of the formed recourse path (Figure\ref{fig:caseStudy}d). This interactive process aids users in understanding the current state across all features~\textbf{(T1)}.

In this view, hovering over dots enables direct comparison with the last selected point, supporting comparisons across multiple views to help users determine the next state~\textbf{(T5)} and form effective recourse paths~\textbf{(T6)}. For example, when a user selects their data point on the Recourse Path Explorer, \systemname instantly displays the associated outcome and feature attributions, with large bubbles indicating actionable next steps~\textbf{(T4)}; hovering over these bubbles further highlights differences in feature values and attributions between the current and potential next states~\textbf{(T1}, \textbf{T5)}.


\par \noindent  \textbf{Outcome Monitor}: Once a dot is selected on the Recourse Path Explorer, the Outcome Monitor displays a modified connected scatter plot paired with a modified deviation plot. In the upper half design, the x-axis represents the recourse path steps while the y-axis depicts the model outcome and its feature attributions~(Figure~\ref{fig:caseStudy}b). For each state, negative attributions are stacked downward from a zero baseline, and positive attributions are built upward from their lowest point, with the endpoint marked by the human icon showing the resulting model outcome. The lower half maintains recourse steps on the x-axis with individual features on the y-axis~(Figure\ref{fig:caseStudy}c), using dotted lines to represent normalized feature ranges, black dots for average values, and colored dots for current values—connected by solid lines to visualize deviations from averages.
Repeated for each recourse state, this visual analytic design allows users to gauge features' impact on outcomes~\textbf{(T2)} and assess goal proximity by comparing the positive attribution stack endpoints with target outcomes. The modified deviation plot further reveals which features lie significantly above or below the norm—indicating areas for refinement~\textbf{(T3)}—and together, these visualizations provide an overall view of progress along the recourse path and the adjustments needed to meet the target outcome.

\subsubsection{Color and Interactivity}
\systemname uses a discrete set of colors to visualize six key features and the recourse path, following explainable AI practices while grouping the rest as ``others." Since SHAP is additive, this subset approach doesn't violate constraints and reduces cognitive load while allowing users to select any features for path construction. Though features beyond the top six are excluded from the Recourse Path Explorer, their collective influence appears in gray on the Outcome Monitor, ensuring awareness of their impact.
When users select a dot, \systemname initiates a sequence of data operations where a Flask-based backend identifies potential targets as per defined constraints, recalculates SHAP attributions, and returns data to adjust dot coordinates with animated transitions as dots reposition and resize. Through repeated selections, a recourse path forms across views with special encoding for starting points (unfilled circles) and endpoints (filled circles), while deselection enables backtracking. This coordinated interactivity between views facilitates effective incremental planning by allowing exploration of alternative paths.
\section{Usage Scenario: Credit Risk Recourse Planning}
        

We present a usage scenario derived from pilot studies to demonstrate how visual analytics could support incremental recourse planning. Let us consider a scenario where, Allen, a business professional, is interested in understanding her chances for credit approval.
Allen began by selecting the Credit Risk dataset~\cite{creditRisk} and examining feature values and attributions~\textbf{(T1)}. She identified a dot representing income at ~\$70k with higher attribution—a counterintuitive finding since lower income generally correlates with lower loan probability~(Figure~\ref{fig:caseStudy}a). In the Outcome Monitor, she saw that despite positive income attribution, other features negatively influenced the loan decision~\textbf{(T2)} (Figure~\ref{fig:caseStudy}b). In the lower half, she noted that most features were near average, but loan interest rate and loan amount were well below average~\textbf{(T3)} (Figure~\ref{fig:caseStudy}c). This aligns with expectations that lower income combined with less favorable interest rates reduces approval chances, as higher-income applicants typically secure better loan terms.

After achieving the initial goal of understanding the current state through feature attributions and the model outcome, Allen proceeded to identify potential targets for the next state. \systemname automatically suggested three potential targets, depicted as larger dots~\textbf{(T4)}, and she hovered over these and other candidates to assess their distance from the current state and check for increased attribution across features~\textbf{(T5)}, paying close attention to steeper slopes indicating small feature changes yielding large positive attributions given their real-life impact. When a selected state didn't significantly improve the model outcome, Allen used the `Undo' button to revert; through several iterations, she eventually identified a recourse path yielding substantial growth in the model outcome~($\sim$80\% approval odds)~(Figure~\ref{fig:caseStudy}b). Thus, \systemname facilitated comparing different recourse paths, enabling selection of one with maximum outcome growth and minimal changes to features like income, employment length, and credit history length— which are generally within an applicant's capacity~\textbf{(T6)}.

While analyzing the recourse path, Allen noted that certain features like income and employment length exhibited partial downward trajectories despite outcome improvements, but loan interest rate, loan amount, and credit history length displayed upward trajectories~(Figure~\ref{fig:caseStudy}d)-contributing to high model outcomes~(Figure~\ref{fig:caseStudy}b). Although credit history length and employment length showed large attribution increases from the initial state, their positive impact was smaller than loan interest rate and loan amount in the final recourse state~(Figure~\ref{fig:caseStudy}b), suggesting feature importance varies across different recourse stages. The significant outcome improvement was driven by increased loan interest rates and decreased loan percent income, aligning with model explanations and confirming \systemname's visual encodings faithfully represent these relationships. 


\section{Subjective User Feedback}

To evaluate the faithfulness of \systemname's visual encodings to SHAP attributions and its effectiveness in guiding recourse path generation for users, we conducted observational studies with $12$ graduate students who had a background in computation and varied levels of familiarity with explainable AI techniques. Because of the inherent individual differences in user preferences for recourse planning~\cite{lam2011empirical}, designing metrics for evaluating accuracy or efficiency is challenging. Therefore, the scope of our evaluation was anchored on assessing the subjective user experience~\cite{koh2025understanding} with a focus on observing if \systemname users can reliably engage in recourse-relevant decision-making tasks.
Each session lasted approximately $45$-$60$ minutes, during which participants completed recourse planning tasks while verbalizing their thought processes. From the feedback collected, we identified several key themes: \textbf{i) Impact of Visualizations:} Participants consistently highlighted the system's ability to reveal feature relationships and their impact on outcomes, demonstrating how visual analytics can bridge the gap between complex model behavior and human-interpretable patterns. One participant noted, ``\textit{The interface is very useful in seeing the impact of different attributes on an end result. It was very well-organized and interactions were clear}," while another commented, ``\textit{Very neat. Provides a fast visualization to help with planning.}" The visual encodings were particularly appreciated, with one participant stating, ``\textit{Visualizations are interactive and well represented.}" \textbf{ii) Easy to Use:} Participants recognized the system's adaptability across different recourse scenarios, reinforcing the importance of accommodating diverse analytical approaches in visual analytics systems. ``\textit{Cool interface for analyzing/recoursing for any purpose in general}," observed one participant, while another remarked, ``\textit{It's visually appealing, and I like that the results are based on real students' data}," \textbf{iii) Advantages over existing approaches:} Participants who had prior experience with making recourse decisions appreciated the incremental flavor of recourse planning in \systemname as compared to traditional trial-and-error approaches. One mentioned,  ``\textit{I found the ReVise interface to be very interesting... I did see it as very valuable for helping someone make choices regarding how to improve aspects such as grades or GPA}"- highlighting the value of data-driven recommendations over guesswork. \textbf{iv) Scope for Improvement:} The primary critique centered on the initial learning curve. One participant shared, ``\textit{At first glance, it may not look useful, and I felt a bit confused about how it was going to help me.}" Another noted difficulty in understanding ``\textit{how the features correlated and change}," attributing this to ``\textit{having little knowledge about the background}." However, several participants also highlighted the value of embedded guidance, with one noting, ``\textit{After I read the helper texts, everything became more clear, and I found those instructions the most useful ones.}" 

\section{Conclusion}
In this work, we propose \systemname, an interactive visual analytic interface for incremental recourse planning that helps users navigate and select effective recourse paths to achieve favorable outcomes from black-box machine learning models. This work advances the state of algorithmic recourse by shifting the focus from deterministic outcomes to probability-aware incremental paths, thereby providing users with greater agency when navigating the increasingly complex algorithmic systems that influence their lives. For future work, we will extend the interface to support recourse planning for groups with different strategies~\cite{fonseca2023setting}, handle incremental model shifts, analyze recourse robustness~\cite{upadhyayRobustReliableAlgorithmic2021}, enable flexible feature selection and projection, elicit user preferences, and incorporate time-span constraints for recourse steps.

\section*{Supplemental Materials}
\label{sec:supplemental_materials}
A demonstration video is available at \url{https://youtu.be/PkpE7ZHoccQ}.

\acknowledgments{
This work was funded in part by the National Science Foundation (NSF) grants 2312932 and 2326195. This paper is released as PNNL-SA-212903.}

\bibliographystyle{abbrv-doi}

\bibliography{bib}
\end{document}